\documentclass[a4paper]{article}
\usepackage{graphics,hyperref}
\usepackage{hyperref}
\usepackage{amsmath}
\usepackage[english]{babel}
\usepackage{latexsym}
\usepackage{amssymb}
\usepackage{amscd}
\usepackage{amsgen,amstext,amsbsy,amsopn}
\usepackage{math rsfs}

\usepackage{amsthm,epsfig,graphicx,graphics}
\usepackage[latin1]{inputenc}
\usepackage{xspace}
\usepackage{amsxtra}
\usepackage{bm}

\usepackage{pdfsync}
\usepackage[english]{babel}
\usepackage{latexsym}

\usepackage{amsmath}

\usepackage{amsfonts}

\usepackage{amssymb}
\usepackage{amscd}
\usepackage{amsgen,amstext,amsbsy,amsopn}
\usepackage{math rsfs}

\newcommand{\bdm}{\begin{displaymath}}
\newcommand{\edm}{\end{displaymath}}
\newcommand{\bdn}{\begin{eqnarray}}
\newcommand{\edn}{\end{eqnarray}}
\newcommand{\bay}{\begin{array}{c}}
\newcommand{\eay}{\end{array}}
\newcommand{\ben}{\begin{enumerate}}
\newcommand{\een}{\end{enumerate}}
\newcommand{\beq}{\begin{equation}}
\newcommand{\eeq}{\end{equation}}
\newcommand{\beqq}{\begin{equation*}}
\newcommand{\eeqq}{\end{equation*}}

\newcommand{\lf}{\left}
\newcommand{\ri}{\right}

\newcommand{\R}{\mathbb{R}}

\newcommand{\xv}{\mathbf{x}}

\newcommand{\rv}{\mathbf{r}}

\newcommand{\diff}{\mathrm d}
\newcommand{\eps}{\varepsilon}

\newcommand{\E}{\mathcal{E}}

\newcommand{\gpf}{\mathcal{E}^{\mathrm{GP}}}

\newcommand{\magnp}{\mathbf{A}}

\newcommand{\half}{\frac12}

\newcommand{\Ofirst}{\Omega_{\mathrm{c_1}}}
\newcommand{\Osec}{\Omega_{\mathrm{c_2}}}
\newcommand{\Othird}{\Omega_{\mathrm{c_3}}}

\newcommand{\gvf}{\mathcal E_{\mathrm{ gv}}}
\newcommand{\gve}{E_{\mathrm{ gv}}}
\newcommand{\gvm}{g_{\mathrm{ gv}}}

\newcommand{\tx}{\textstyle}
\newcommand{\OO}{\mathcal{O}}

\begin{document}
\title{Giant vortex phase transition in rapidly rotating trapped Bose-Einstein condensates}
\author{M. Correggi${}^{a}$, F. Pinsker${}^{b}$, N. Rougerie${}^{c}$, J. Yngvason${}^{d,e}$\\
	\mbox{}	\\
	\normalsize\it ${}^{a}$ Dipartimento di Matematica, Universit\`{a} degli Studi Roma Tre,	\\
	\normalsize\it L.go S. Leonardo Murialdo 1, 00146, Rome, Italy.	\\
	\normalsize\it ${}^{b}$ DAMTP, University of Cambridge, Wilbertforce Road, Cambridge CB3 0WA, United Kingdom.\\
	\normalsize\it ${}^{c}$ Universit\'e Grenoble 1 and CNRS, LPMMC, UMR 5493, BP 166, 38042 Grenoble, France.\\
	\normalsize\it ${}^{d}$ Fakult\"at f\"ur Physik, Universit{\"a}t Wien, Boltzmanngasse 5, 1090 Vienna, Austria.	\\
	\normalsize\it ${}^{e}$ Erwin Schr{\"o}dinger Institute for Mathematical Physics, Boltzmanngasse 9, 1090 Vienna, Austria.}	

\date{October 26th, 2012}

\maketitle

\begin{abstract}A Bose-Einstein condensate of cold atoms is a superfluid and thus responds to rotation of its container by the nucleation of quantized vortices. If the trapping potential is sufficiently strong, there is no theoretical limit to the rotation frequency one can impose to the fluid, and several phase transitions characterized by the number and distribution of vortices occur when it is increased from $0$ to $\infty$. In this note we focus on a regime of very large rotation velocity where vortices disappear from the bulk of the fluid, gathering in a central hole of low matter density induced by the centrifugal force. 
\end{abstract}

\section{Introduction}
\label{intro}

\subsection{The model}\label{sec:model}

The main theoretical framework for the description of rotating BECs is the so-called Gross-Pitaevskii (GP) theory. The behavior of the fully Bose-condensed gas is described by a single macroscopic wave function $\Psi$ and the ground state of the condensate is determined by the minimization of the GP energy functional. Assuming a strong confinement along the rotation axis so that the condensate is quasi-2D, the GP functional can be expressed, in the rotating frame, as
\begin{equation}
	\label{GPfunctional}
	\gpf_{\rm phys}[\Psi] = \int_{\mathbb R^2} \lf\{ \half\lf|\lf( {\rm i} \nabla +\magnp \ri) \Psi \ri|^2 + \lf(V_{\rm trap}(\rv)- \half \Omega_{\rm rot}^2 r^2 \ri) |\Psi|^2 + \frac{|\Psi|^4}{\eps^2} \ri\} {\diff} \rv.
\end{equation}
Here  $\magnp=\Omega_{\rm rot}\mathbf{e}_3 \wedge \rv$ with $\Omega_{\rm rot}>0$  the rotation velocity, $ \mathbf e_3$ the unit vector in the $x_3$-direction, $\rv=(x_1,x_2)$, $r=|\rv|$,  $V_{\rm trap}$ the trap potential and  $1/\eps^2$  with $\eps>0$ the GP coupling constant. We work in units such that $\hbar$ and the particle mass are both 1 and the wave-function is normalized as $\int|\Psi|^2=1 $. 

In \eqref{GPfunctional} we have introduced a `vector potential' $ {\bf A} $ in order to separate the contribution of Coriolis and centrifugal forces due to the transformation to the rotating frame. This emphasizes the importance of the \emph{effective potential}
\begin{equation}\label{eq:pot eff}
V_{\rm eff} (\rv)= V_{\rm trap} (\rv)- \half \Omega_{\rm rot}^2 r^2,
\end{equation}
and makes it immediately clear that the physical behavior of the gas will be strongly affected by the type of trap that is being used. In this note we consider a potential of the form 
\beq\label{pot} V_{\rm trap}(r)=kr^s+\half\Omega_{\rm osc}^2r^2\eeq
with $k > 0$ and $s>2$ ({\it anharmonic traps}). Indeed in the purely harmonic (quadratic) case $ k= 0 $, the rotation frequency would be limited by the frequency $\Omega_{\rm osc}$ of the trap and rapid rotation regimes forbidden: Any larger rotation speed would make $V_{\rm eff}$ unbounded from below, and thus the atoms would fly apart under the action of the centrifugal force. If $ k > 0 $ on the other hand, the effective potential \eqref{eq:pot eff} is bounded below for any $\Omega_{\rm rot}$, and there is thus no theoretical obstruction to the exploration of regimes of very large rotation frequencies. 

It is also worth mentioning here that among anharmonic traps ($ s > 2 $) one has to distinguish between `soft potentials' ($ s< \infty $) discussed here and `hard potentials' or `flat traps' considered in \cite{CPRY1,CRY,R} (formally given by $ s = \infty $). Indeed, the physics of the transition to a giant vortex state is different in these two cases and the third critical speed is considerably smaller in the latter (after an appropriate scaling, see below).



\subsection{Phase transitions and critical speeds}\label{sec:crit speed}

Let us briefly review what is expected to happen when the rotation velocity imposed to a trapped Bose gas is increased. We only discuss the case of anharmonic traps ($k>0$) and refer to the literature (see e.g. \cite{Co,Fe2} for reviews) for the specificities of the purely harmonic case.
\begin{itemize}
\item As long as the rotation speed is smaller than a first critical speed $\Ofirst$, the fluid is vortex-free.
\item When $\Omega_{\rm rot}$ is increased past $\Ofirst$, vortices start to appear and form regular patterns. When many vortices are nucleated they are distributed according to a triangular lattice. This behavior lasts until a second critical speed $\Osec$ is attained. 
\item At $\Omega_{\rm rot} \approx \Osec$, the centrifugal force becomes so important that it dips a hole of strongly depleted matter density at the center of the trap. A vortex lattice survives in the annular bulk of the fluid until a third critical speed is reached.
\item When $\Omega_{\rm rot}$ crosses the third critical speed $\Othird$, all the vortices retreat in the central hole of low matter density. The bulk of the fluid is then vortex-free but a phase circulation remains around the hole, indicating that it acts as a multiply quantized \emph{giant vortex}.
\end{itemize}

Experiments using anharmonic traps we are aware of have so far been limited to rotation frequencies $\Omega_{\rm rot} < \Osec$: In \cite{BSSD} a clear density dip is observed at the center of the trap but experimental limitations have prevented the creation of the giant vortex. The theoretical problem is nevertheless of interest and has been intensely investigated \cite{Fe,FJS,FB,FZ,KTU,KB,KF}.

In this note we focus on the third phase transition, when the state of the condensate changes from a ``vortex-lattice-plus-hole'' to a ``giant vortex''. We explain the main physical insights leading to a mathematically rigorous estimate of the critical speed we have recently obtained. A more complete discussion and the details of the proof can be found in \cite{CPRY2,CPRY3}, together with a more thorough discussion of the literature.

\section{Emergence of the giant vortex state}\label{sec:giant vortex}

\subsection{Scaling of the GP functional}\label{sec:scale}

In the rest of this note we will consider the strongly interacting (Thomas-Fermi) regime $\eps \to 0$ and focus on angular velocities $ \Omega_{\rm rot} \gg 1 $ close to the third critical speed.  

In order to make the effect of the quadratic part in the potential $  V_{\rm trap} $ apparent even in the limit $\Omega_{\rm rot}\to\infty $, we keep the ratio $\Omega_{\rm eff}/\Omega_{\rm rot}$ fixed by setting $\Omega_{\rm eff}^2=\gamma\Omega_{\rm rot}^2$ for some $0<\gamma\leq 1$ fixed, where $\Omega_{\rm eff}=(\Omega_{\rm rot}^2-\Omega_{\rm osc}^2)^{1/2}$, so that the effective potential becomes 
$V_{\rm eff} (r) = kr^s- \half \Omega_{\rm eff}^2 r^2$.

Moreover we scale variables in such a way that the bulk of the mass of the condensate does not expand as $ \Omega_{\rm rot} \to \infty $: Most of the mass is indeed concentrated close to the minimum of the potential $ V_{\rm eff} (r) $ at $r=R_{\rm m}$ with $ R_{\rm m}\label{Rm}= [\Omega_{\rm eff}^2/(sk)]^{1/(s-2)} $, 
which wanders to infinity as $\Omega_{\rm rot}\to\infty $. It is thus convenient to define the scaling in such a way that the global minimum of the rescaled potential is independent of the angular velocity. This can be done by introducing the following variables
\beq\label{scaling}
\mathbf x = R_{\rm m} ^{-1} \rv, 
\quad \psi(\mathbf x) = R_{\rm m} \Psi(\mathbf r),\quad \Omega = R_{\rm m}^{2} \Omega_{\rm rot} \eeq 
which yields $\mathcal E^{\rm GP}_{\rm phys}[\Psi]=R_{\rm m}^{-2}\mathcal E^{\rm GP}[\psi]$ with the scaled energy functional
\begin{equation}
\label{gpscaled}\mathcal E^{\rm GP}[\psi]=\int_{\mathbb R^2}\left\{\half|(\mathrm i\nabla+\mathrm \Omega x\mathbf e_{\vartheta})\psi|^2+\gamma\Omega^2V(x)|\psi|^2+\eps^{-2}|\psi|^4\right\}\diff\mathbf x 
\end{equation}
and the potential (with minimum at $ x = 1 $) 
\beq\label{effpot2} V(x) = \frac1{s} x^s-\half x^2 .\eeq

The potential and interaction terms in \eqref{gpscaled} become comparable when $\Omega \sim \eps^{-1}$, which is the order of the  second critical speed $\Osec$ (see \cite{CPRY2,CPRY3}). In the sequel we take $\Omega \gg \eps ^{-1}$ (and will in fact specialize to $ \Omega \gtrsim \eps^{-4} $) and investigate the features of the transition associated with the third critical speed. 


\subsection{The matter density profile}\label{sec:profile}

As a first step we look for a good approximation to the density profile of the fluid. To this end we define the functional
\begin{equation}\label{giantfunc}
\mathcal \gvf [g] =\int_{\mathbb R^2}\left\{\half|\nabla g|^2+\half\Omega^2(x-x^{-1})^2g^2 +\gamma\Omega^2\left(\frac{1}{s}x^s-\frac{1}{2} x^2\right)g^2+\eps^{-2}g^4\right\}\diff{\mathbf x}
.
\end{equation}
This is nothing but \eqref{gpscaled} restricted to wave functions of the form\footnote{Strictly speaking we should use the integer part of $\Omega$ in the phase factor, but since $\Omega$ is very large in our regime, it makes little difference to assume that it is an integer.}
\beq\label{ansatz} \psi(\xv)= g(\xv)\exp(\mathrm i\Omega\vartheta)\eeq
with a real valued function $g$, normalized such that $\int g^2=1$. In the giant vortex ansatz \eqref{ansatz}  all the vorticity is concentrated in a single vortex at the origin with huge winding number $ \Omega $. We denote $\gvm$ and $\gve$ the ground state and ground state energy of $\gvf$ respectively.

A noticeable fact is that when $\Omega = \Omega_0 \eps ^{-4}$ with $\Omega_0 = \OO(1)$, $\gvm$ progressively changes from a Thomas-Fermi to a gaussian profile. This is most conveniently understood by a change of variables. Since the minimizer of \eqref{giantfunc} is radial we consider only radial functions  and define $y=\Omega^{1/2}(x-1)$, and $ f(y)=\Omega^{-1/4}g(x)$. Equation \eqref{giantfunc} then turns into
\begin{equation}\label{auxfunc}
{\mathcal E}^{\mathrm{aux}}[f] = \Omega\int_{\mathbb \R } \left\{\half | f'|^2+\half \alpha^2y^2  f^2+\Omega_0^{-1/2} f^4\right\}{\diff} {y}
\end{equation}
up to an unimportant constant term. We have set $\alpha^2=4+\gamma(s-2)$, Taylor-expanded the potential and neglected terms beyond quadratic (see \cite[Section III.B]{CPRY2}). Without the last term the minimizer is the gaussian
\beq f_{\rm osc}(y)=\pi^{-1/4}\alpha^{1/4}\exp\lf\{-\tx{\half} \alpha y^2\ri\}\eeq
which will give a good approximation (after scaling) to the giant vortex profile $\gvm$ when $\Omega_0$ is large, i.e., $\Omega \gg \eps ^{-4}$. Notice that it is quite different, in particular because of its long tails, from the TF profile (with $\mu_{\rm TF}$ a chemical potential fixed by normalization and $ [ \: \cdot \: ]_+ $ the positive part)
\beq f_{\rm TF}(y) = \Omega_0 ^{1/4} \sqrt{\left[ \mu_{\rm TF} - \tx\frac{1}{4}\alpha ^2 y ^2 \right]_+}, \eeq
which is a better approximation of $ \gvm $ for $ \Omega \ll \eps^{-4} $.

An important step of our approach is an \emph{energy decoupling}: Defining, for any normalized wave function $\psi$ a function $v$ by writing $ \psi = \gvm  v \exp(\mathrm i\Omega\vartheta)$ and using the variational equation satisfied by $\gvm$, we get
$\label{eq:decoupling} \gpf[\psi] = E^{\rm gv} + \mathcal E[v]$
with
\beq\label{eeta} 
\mathcal E[v]=\int_{\R ^2} \gvm^2\left\{\half|\nabla v|^2-\mathbf B\cdot \mathbf J(v)+\eps^{-2} \gvm ^2(1-|v|^2)^2\right\} \diff \xv.\eeq
We denote by $\mathbf J(v)= \tx\frac{i}{2}(v\nabla v ^* - v^* \nabla v )$ the superfluid current associated to $v$ and $\mathbf B= \Omega(x-x^{-1})\mathbf e_\vartheta$ the vector potential corrected by the contribution of the giant vortex.

Our analysis of the phase transition leading to the giant vortex state is based on the minimization of the reduced functional \eqref{eeta} under the constraint $\int \gvm ^2 |v| ^2 = 1$. Thanks to the energy decoupling this is equivalent to the original variational problem: A function $u$ minimizes \eqref{eeta} if and only if $\psi = \gvm u \exp(\mathrm i\Omega\vartheta)$ minimizes the scaled GP functional \eqref{gpscaled}. Note that $\gvm$ does not vanish, except at the origin, and thus all the potential vortices of $\psi$ are accounted for by the reduced wave function $u$.

\subsection{Disappearance of vortices}\label{sec:vortices}

Now that we have a good approximation to the matter density profile, we can discuss the nucleation of vortices in the bulk of the fluid, that is in the region where $\gvm$ is significantly large. We first discuss the size that vortex candidates would have. Using an ansatz $v$ that vanishes inside the bulk in a region (vortex core) of size $t$ and minimizing \eqref{giantfunc} with respect to $t$ gives an estimate of the size of vortex cores : $t\sim \eps^{2/3}\Omega^{-1/3}$. From the considerations in the above section, one can guess that the thickness of the annulus where the mass resides is of order $\Omega^{-1/2}$. 
When $\Omega \gtrsim \eps ^{-4}$ we have $t\gtrsim \Omega ^{-1/2}$ which means that vortices become too large to completely fit in the bulk of the fluid, whereas they are relatively small when $\Omega \ll \eps^{-4}$. This alone is of course no proof that a phase transition occurs in this regime but it turns out that one can, when $\Omega \ll \eps ^{-4}$, rigorously estimate the energetic contribution of vortices and prove that many are nucleated and uniformly distributed in the bulk. This result is obtained by making use of the powerful ``vortex balls methods'' (see, e.g., \cite{A,SS} for reviews) that apply in this regime. 

The regime $\Omega \gtrsim \eps ^{-4}$ on the other hand poses a new challenge. Since vortex candidates are comparable in size to the bulk of the fluid, the intuition behind the vortex balls methods fails and a different approach has to be used. In fact, one can rule out density depletions without any precise estimate of the vortices' energy. The argument goes as follows\footnote{With significant simplifications, see \cite{CPRY2,CPRY3} for the full details.}: a trial state $v = 1$ indicates that if $u$ minimizes \eqref{eeta}, then $\mathcal{E} [u] \leq 0$. If on the other hand we can prove that $\E[u] \gtrsim 0$ with a sufficiently small error, it is clear that it is not favorable to nucleate vortices, since the trial state $v = 1$ is then almost optimal.

To estimate the energy it is convenient to integrate the second term of \eqref{eeta} by parts : 
Since $\nabla\cdot \gvm ^2 \mathbf B= 0$, we can write $\gvm ^2 \mathbf B=\nabla^\perp F$ with the dual gradient $\nabla^\perp=(-\partial_{x_2},\partial_{x_1})$ and a suitable potential function $F$. By Stokes' formula we then have\footnote{In full rigor, the integration should be restricted to a bounded region and boundary terms included in the analysis.}
\beq\label{intj} 
-\int_{\R ^2} \gvm ^2\, \mathbf B\cdot \mathbf J(u)=\int_{\R ^2} F\,\nabla^\perp\cdot\mathbf J(u).
\eeq
Now, it is clear that the only potentially negative term in \eqref{eeta} is \eqref{intj}. It is thus the only one that could favor the nucleation of vortices. Note also that the quantity $\nabla^\perp\cdot\mathbf J(u)$ appearing in the r.h.s. of \eqref{intj} is nothing but the $\mathrm{curl}$ of the superfluid current. It can thus (in analogy with fluid mechanics) be thought of as a vorticity measure. A trivial estimate yields $\left|\nabla^\perp\cdot\mathbf J(u)\right| \leq |\nabla u| ^2$ and we thus have
\begin{equation}\label{pouet}
 \int_{\R ^2} \half \gvm ^2 |\nabla u| ^2 -\int_{\R ^2} \gvm ^2\, \mathbf B\cdot \mathbf J(u) \geq \int_{\R ^2} \left( \half  \gvm ^2 - |F| \right) |\nabla u| ^2.
\end{equation}
The potential $F$ is an explicit function of $\gvm$ and, using for the latter the gaussian ansatz described in the preceding section, one can see that $2|F|\leq \gvm ^2$. When $\Omega_0$ is sufficiently large for the gaussian approximation to become accurate one can then infer that \eqref{pouet} is $\gtrsim 0$, which indicates that nucleating vortices can not decrease the energy. This is the desired lower bound to the energy $\E[u]$ and actually gives more information: Since we have not used the third term in \eqref{eeta} so far, we deduce an upper bound to this term. A careful analysis then reveals that energy considerations prevent $u$ from vanishing in the bulk of the condensate.

As mentioned before the behavior in the case of hard trapping potentials ($ s = \infty $) is quite different: the squeezing of the condensate against the hard walls of the trap due to the centrifugal force is much more pronounced is this case and vortices therefore energetically more costly. This explains why the third
critical speed is considerably larger for soft trapping potentials ($ \Othird \gg \eps^{-4} $) than for hard potentials ($ \Othird \sim \eps^{-2} |\log\eps|^{-1} $). 

\section{Conclusions}\label{sec:conclusions}

We have sketched the energetic considerations leading to our proof that a phase transition from a vortex lattice to a giant vortex state occurs in BECs rotated at a very large rotation speed. Our main finding is that in the rescaled variables introduced in Section \ref{sec:scale}, the transition happens when $\Omega  \gg \eps ^{-4}$, i.e., $ \Omega_{\rm rot} \gg \eps^{-4(s-2)/(s+2)} $ in the physical variables. It would be interesting to know a precise estimate of the critical speed for the phase transition, but since the potential vortex cores are comparable in size to the bulk of the fluid, it is not even clear that a sharp transition, rather than a smooth crossover, occurs. 

\medskip
\textbf{Acknowledgements.}\\
M.C. gratefully acknowledges financial support from the European Research Council under the European Union's Seventh Framework Programme ERC Starting Grant CoMBoS (grant agreement n. 239694). N.R. was partially supported by the same programme (grant agreement MNIQS-258023).

\end{document}